\documentclass[12pt]{iopart}

\usepackage{iopams,amssymb,epsfig,subfigure,graphicx,amsthm}

\usepackage{setstack}
\usepackage{amscd}
\usepackage{amsthm}
\usepackage{amssymb}
\usepackage{amsbsy}
\usepackage{makeidx}
\usepackage{epsf}
\usepackage{graphics}
\usepackage{graphicx}
\usepackage{lscape}
\usepackage{amsfonts}
\usepackage{color}
\usepackage{subfigure}

\newcommand{\sech}{\,\mathrm{sech}}

\begin{document}

\title[Fluxon analogues and dark solitons in linearly coupled BECs]{%
Fluxon analogues and dark solitons in linearly coupled Bose-Einstein condensates}
\author{M.I.\ Qadir$^{1,2}$, H.\ Susanto$^1$ and P.C.\ Matthews$^1$}
\address{$^1$ School of Mathematical Sciences, University of Nottingham,\\
University Park, Nottingham NG7 2RD, UK}
\address{$^2$ Department of Mathematics, University of Engineering and Technology,\\ Lahore, Pakistan}
\ead{hadi.susanto@nottingham.ac.uk}

\begin{abstract}
Two effectively one-dimensional parallel coupled Bose-Einstein condensates in the presence of external potentials are studied. The system is modelled by linearly coupled Gross-Pitaevskii equations. In particular, grey-soliton-like solutions representing analogues of superconducting Josephson fluxons as well as coupled dark solitons are discussed. Theoretical approximations based on variational formulations are derived. It is found that the presence of a magnetic trap can destabilize the fluxon analogues. However, stabilization is possible by controlling the effective linear coupling between the condensates. 
\end{abstract}

\maketitle



\newpage

\section{Introduction}

The concept of electron tunnelling between two superconductors separated by a thin insulating barrier predicted by Josephson \cite{jose62} has been extended relatively recently to tunnelling of Bose-Einstein condensates (BECs) across a potential barrier by Smerzi et al.\ \cite{smer97,ragh99,giov00}. Such tunnelling has been observed experimentally where a single \cite{albi05,levy07} and an array \cite{cata01} of \emph{short} Bose-Josephson junctions (BJJs) were realized. The dynamics of the phase difference between the wavefunctions of the condensates \cite{smer97,ragh99,giov00,ostr00,anan06,jia08} resembles that of point-like Josephson junctions \cite{baro82}.

Recently a proposal for the realization of a \emph{long} BJJ has been presented by Kaurov and Kuklov \cite{kaur05,kaur06}. Similarly to superconducting long Josephson junctions, one may also look for an analogue of Josephson fluxons \cite{usti98} in this case. It was shown in \cite{kaur05,kaur06} that fluxon analogues are given by coupled dark-soliton-like solutions, as the relative phase of the solutions has a kink shape with the topological phase difference equal to $2\pi$. Moreover, it was emphasized that fluxon analogues (FAs) can be spontaneously formed from coupled dark solitons due to the presence of a critical coupling at which the two solitonic structures exchange their stability. The idea of FAs in tunnel-coupled BECs is then extended to rotational FAs in the ground state of rotating annular BECs confined in double-ring traps \cite{bran}.

In this report, we consider the existence and the stability of FAs in two coupled cigar-shaped condensates in the presence of a magnetic trap along the elongated direction 
modelled by the normalized coupled Gross-Pitaevskii equations \cite{kaur05,kaur06}
\begin{equation}
\begin{array}{lll}
\displaystyle i{\psi_j}_t&=&-\frac{1}{2}{\psi_j}_{xx}+\mu|\psi_j|^2\psi_j-\omega\psi_j-k\psi_{3-j}+V\psi_j,\\
\end{array}
\label{nls2}
\end{equation}
where $\psi_j,\,j=1,2,$ is the bosonic field, and $t$ and $x$ are the time and axial coordinate, respectively. 
Here, we assume that the parallel quasi one-dimensional BECs are linked effectively by a weak coupling $k$. Note that herein $k>0$. The case $k<0$ can be obtained accordingly as there is a symmetry transformation $k\to-k$ and $\psi_j\to i\psi_j$. 
The parameters $\mu$ and $\omega$ are respectively the nonlinearity coefficient representing the atomic scattering length and the chemical potential. Here, we consider the defocusing nonlinearity $\mu<0$, see, e.g., \cite{tril88,kivs89,akhm93,dror09}) for the focusing counterpart $\mu>0$. $V$ is the magnetic trap with strength $\Omega$, i.e.\
\begin{equation}
V(x)=\frac{1}{2}\Omega^2 x^2.
\label{ep}
\end{equation}
Extending the work \cite{kaur05,kaur06}, which was without the magnetic trap, i.e.\ $\Omega\equiv0$, we are particularly interested in the effects of $\Omega>0$ to the localized excitations.

When $\Omega=0$, writing $\psi_j=|\psi_j|\exp(i\varphi_j),$ it was shown that the relative phase $\phi=\pm(\varphi_2-\varphi_1)$ will satisfy a modified sine-Gordon equation \cite{kaur05}. A fluxon analogue of (\ref{nls2}) in that case is given by the solution $\psi_1=\psi_2^*=\psi$, with
\begin{equation}
\psi=\pm \sqrt{\frac{\omega+k}{\mu}}\tanh(2\sqrt{k}x)\pm i\sqrt{\frac{\omega-3k}{\mu}}\sech(2\sqrt{k}x),
\label{jv}
\end{equation}
where the asterisk denotes complex conjugation. The soliton (\ref{jv}) can be regarded as an analogue of Josephson fluxons \cite{kaur05,kaur06} as the phase difference $\phi$ between the phases of $\psi_1$ and $\psi_2$ forms a spatial kink connecting $\phi=0$ and $\phi=\pm2\pi$. In the following, solution (\ref{jv}) (and its continuations) will be referred to as FAs. From the expression, it is clear that an FA exists only for $0<k<\omega/3$. For $k=\omega/3$, the solution in (\ref{jv}) transforms into dark solitons \cite{kaur05,kaur06}
\begin{equation}
\psi_{1,2}=\pm \sqrt{\frac{\omega+k}{\mu}}\tanh(\sqrt{\omega+k}x).
\label{ds}
\end{equation}
which exists for $k>-\omega$. Thus solutions in (\ref{jv}) and (\ref{ds}) coincide at $k=\omega/3$ and hence $k=\omega/3$ is the bifurcation point along the family (\ref{ds}). In the following, we denote this critical coupling as $k_{ce}$.


When the two condensates are uncoupled, i.e.\ $k=0$, the dynamics of a dark soliton in BECs with magnetic trap has been considered before theoretically \cite{kono08,fran10} (see also \cite{peli05} and references therein) and experimentally \cite{theo10,burg99,beck08,well08,stel08}. Interesting phenomena on the collective behavior of a quantum degenerate bosonic gas, such as soliton oscillations \cite{beck08,well08,theo10} and frequency shifts due to soliton collisions \cite{stel08} were observed. A theoretical analysis based on variational formulation was developed in \cite{kivs95,kivs98} that is in good agreement with numerics as well as with experiments (see, e.g., \cite{fran02,hong09}). A similar variational method will be derived here to explain the dynamics of FAs.

The present report is outlined as follows. In Section 2 we will derive a relation between the velocity of the travelling FAs and the critical value $k_{ce}$ of the coupling constant at which the FAs solution changes into a coupled dark soliton. In Section 3, a variational formulation for the FAs is considered to study their dynamical behaviours analytically. We will solve the governing equations numerically in Section 4. Comparing the numerical and the analytical results, we show good agreement between them. In the last section, we conclude the work and present open problems for future work.

\section{The velocity dependence of the critical coupling $k_{ce}$}

When FAs do not travel, it has been discussed above that $k_{ce}=\omega/3$. When the solitons move with velocity $v$, it is natural to expect that $k_{ce}=k_{ce}(v)$. It is because dark solitons only exist for $v<1$ \cite{kivs95}, while at $k_{ce}$ FAs become dark solitons. In the following, we are interested in the expression of the critical coupling. We will determine the velocity dependence of $k_{ce}$ analytically.

For simplicity, first we scale the governing equation (\ref{nls2}), such that the nonlinearity coefficient and the chemical potential become $\mu=\omega=1$. We also scale the wavefunction $\psi_j$ in (\ref{nls2}) by $\Psi_j=\psi_j/\sqrt{1+k}$, such that the equation becomes
\begin{equation}
\begin{array}{lll}
\displaystyle i{\Psi_j}_{\tilde{t}}+\frac{1}{2}{\Psi_j}_{\tilde{x}\tilde{x}}-|\Psi_j|^2\Psi_j+\frac{\Psi_j+k\Psi_{3-j}}{1+k}=\frac{V\Psi_j}{1+k},
\end{array}
\label{nls3}
\end{equation}
where $\tilde{t}=(1+k)t$ and $\tilde{x}=\sqrt{1+k}\,x$.

Defining $\tilde{\xi}=\tilde{x}-\tilde{v}\tilde{t}$, Eq.\ (\ref{nls3}) with $V\equiv0$ in a moving coordinate frame can be written as
\begin{equation}
\begin{array}{lll}
\displaystyle -i\tilde{v}{\Psi_j}_{\tilde\xi}+\frac{1}{2}{\Psi_j}_{{\tilde \xi}{\tilde \xi}}-|\Psi_j|^2\Psi_j+\frac{\Psi_j+k\Psi_{3-j}}{1+k}=0.
\end{array}
\label{nls-3}
\end{equation}
The equation above has an explicit solution for the travelling dark soliton \begin{equation}
\psi_{ds}=A\tanh(B\tilde{\xi})+i \tilde{v},
\label{tamb}
\end{equation}
with $A=B$ and $A^2+\tilde{v}^2=1$. 

Let us perturb $\psi_{ds}$ such that
$\psi_j=\psi_{ds}+(-1)^{3-j}\epsilon f(\tilde{\xi})$, where $\epsilon\ll1$. Substituting these values of $\psi_1$ and $\psi_2$ in (\ref{nls-3}) and retaining linear terms in $\epsilon$ only, we obtain
\begin{eqnarray*}
-\frac{1}{2}\frac{d^2f(\tilde{\xi})}{d\tilde{\xi}^2}+\biggr[2(A^2\tanh^2(B\tilde{\xi})+v^2)+k-1\biggr]f(\tilde{\xi}) \\
+\biggr[A^2\tanh^2(B\tilde{\xi})-v^2\biggr]f^*(\tilde{\xi})+i\biggr[2Av\tanh(B\tilde{\xi})f^*(\tilde{\xi})
+v\frac{d f(\tilde{\xi})}{d\tilde{\xi}}\biggr]=0,
\end{eqnarray*}
where we have substituted $v=\sqrt{1+k}\tilde{v}$. Here, $v$ is the velocity measured in the `original' time $t$, while $\tilde{v}$ is in the scaled time $\tilde{t}$.

Choosing
\begin{equation}
f(\tilde{\xi})=p\,\sech(B\tilde{\xi})\tanh(B\tilde{\xi})+i\,q\,\sech(B\tilde{\xi}),
\label{corr}
\end{equation}
where $p$ and $q$ are real, and substitute it in the last equation above, 
we obtain two equations containing $p$ and $q$. Eliminating $p$ and $q$ in the resulting equations, we obtain the following relation between $k$ and $v$
\begin{equation}
k=-\frac{1}{3}{v}^{2}-\frac{1}{21}+\frac{4}{21}\sqrt{7v^4-7v^2+4}.\label{k_v}
\end{equation}
The coupling constant $k$ here is the critical coupling $k_{ce}$, which is clearly a function of $v$. Note from the above expression that $k_{ce}\rightarrow \frac{1}{3}$ as $v\rightarrow 0$ and  $k_{ce}\rightarrow 0$ as $v\rightarrow 1$.

\section{Variational formulations of the FAs}

The Lagrangian formalism for the FAs will be separated into two cases, i.e.\ when the coupling constant $k$ is close to the critical coupling $k_{ce}$ at which the FAs are close to coupled dark solitons and when it is close to the uncoupled limit $k\approx0$. The two cases determine the ansatz that will be taken to approximate the solutions.

\begin{figure*}[tbhp!]
\begin{center}
\subfigure[]{\includegraphics[width=7cm]{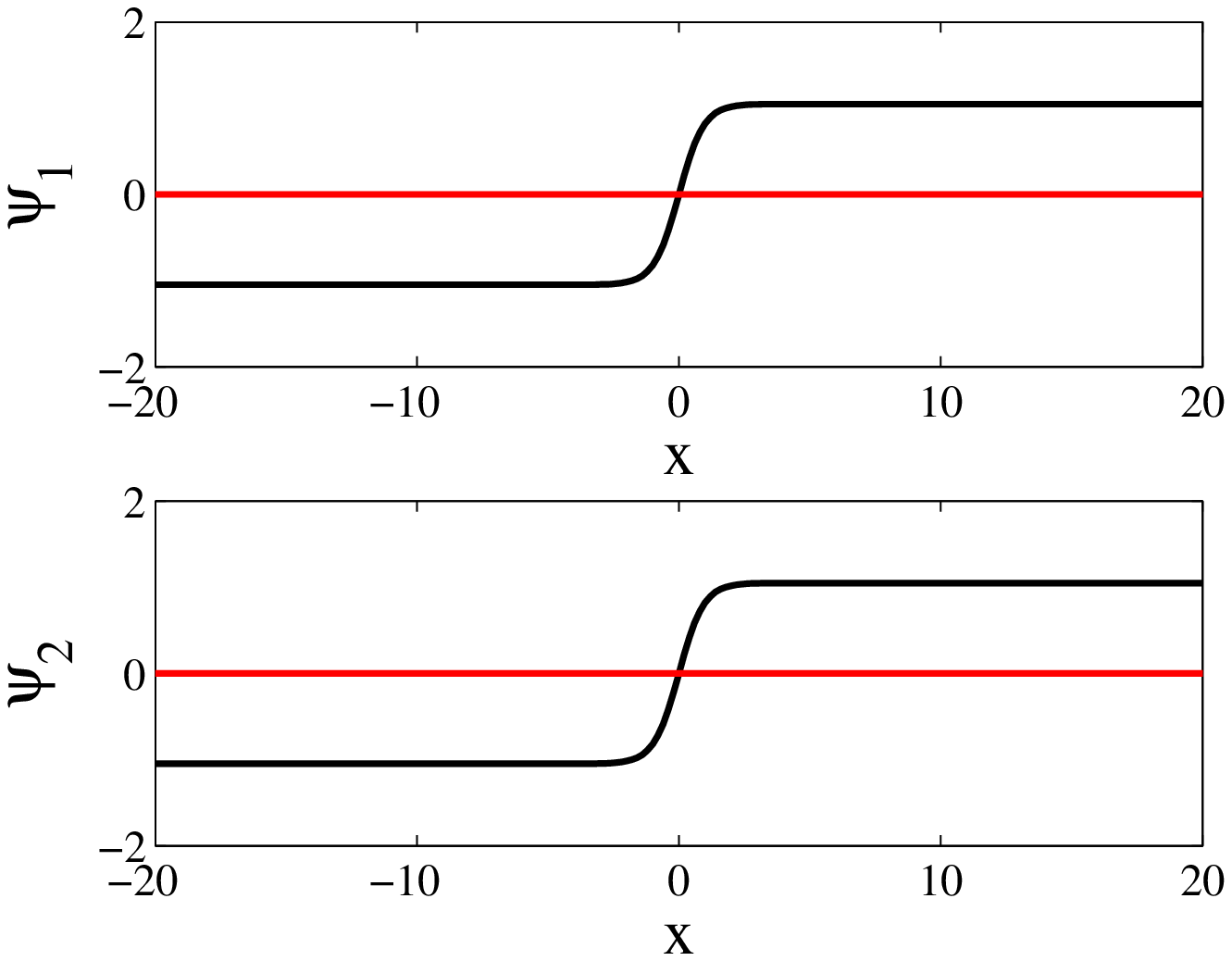}\label{Fig 1}}
\subfigure[]{\includegraphics[width=7cm]{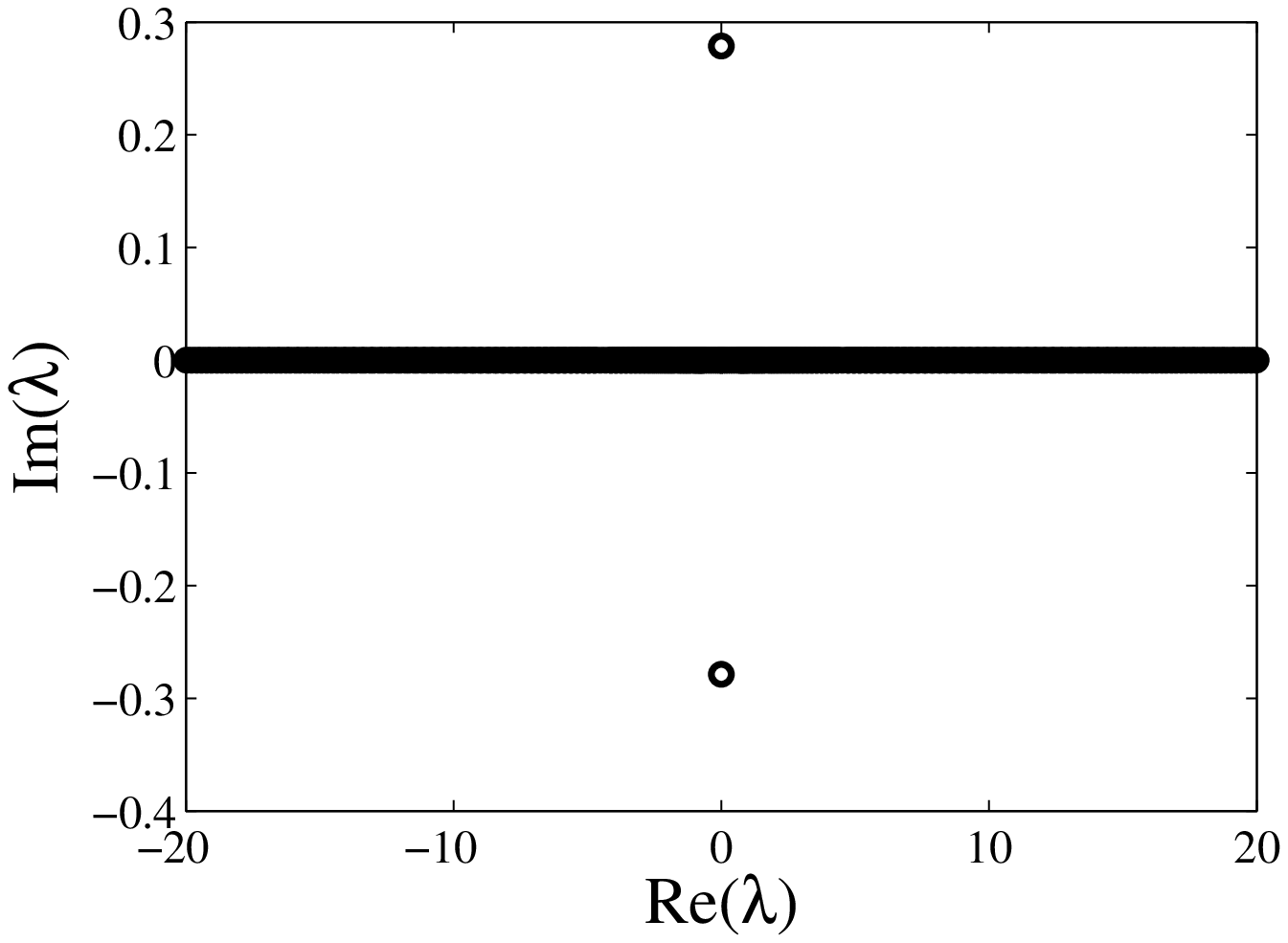}\label{Fig 4}}
\subfigure[]{\includegraphics[width=7cm]{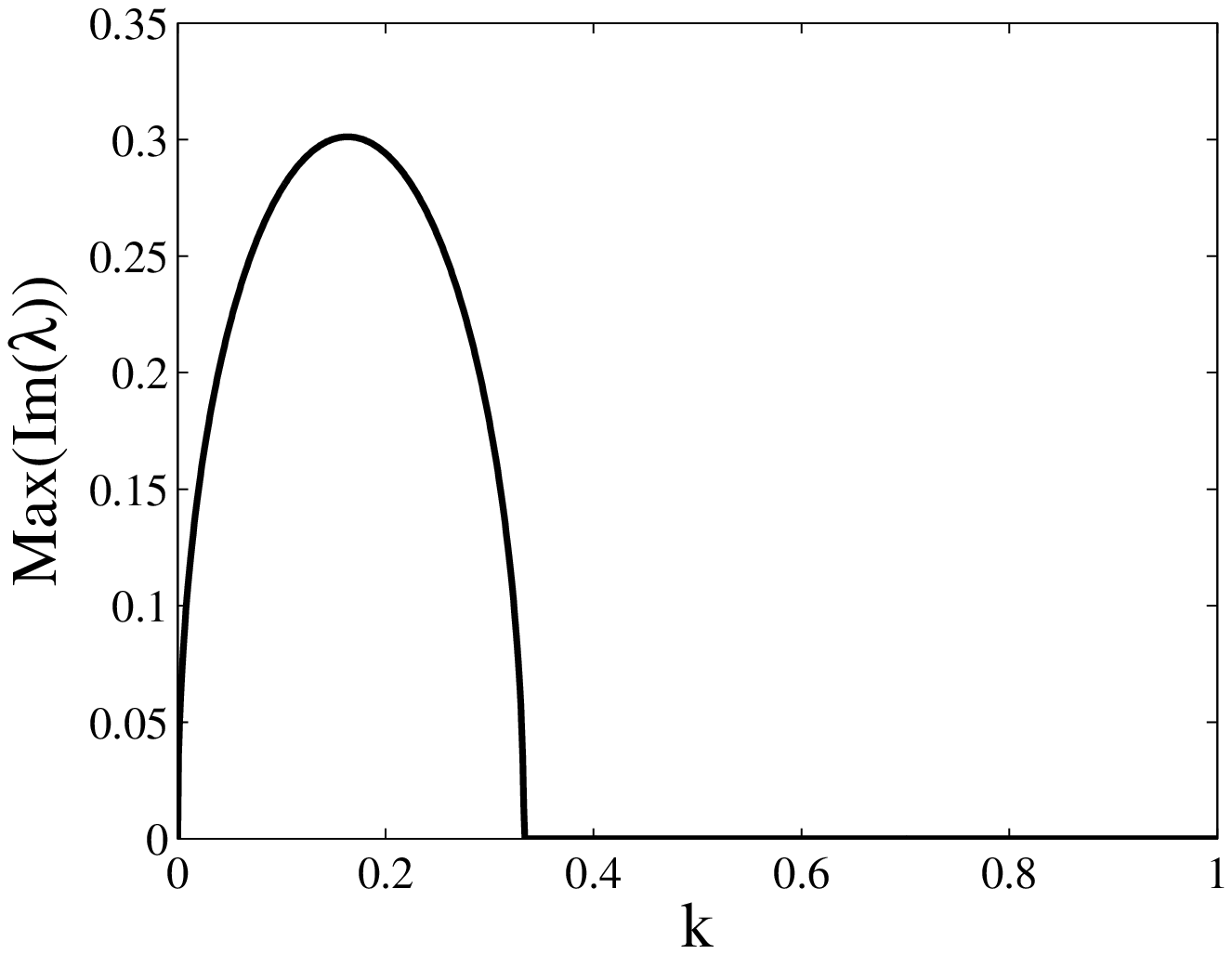}\label{Fig 2}}
\subfigure[]{\includegraphics[width=7cm]{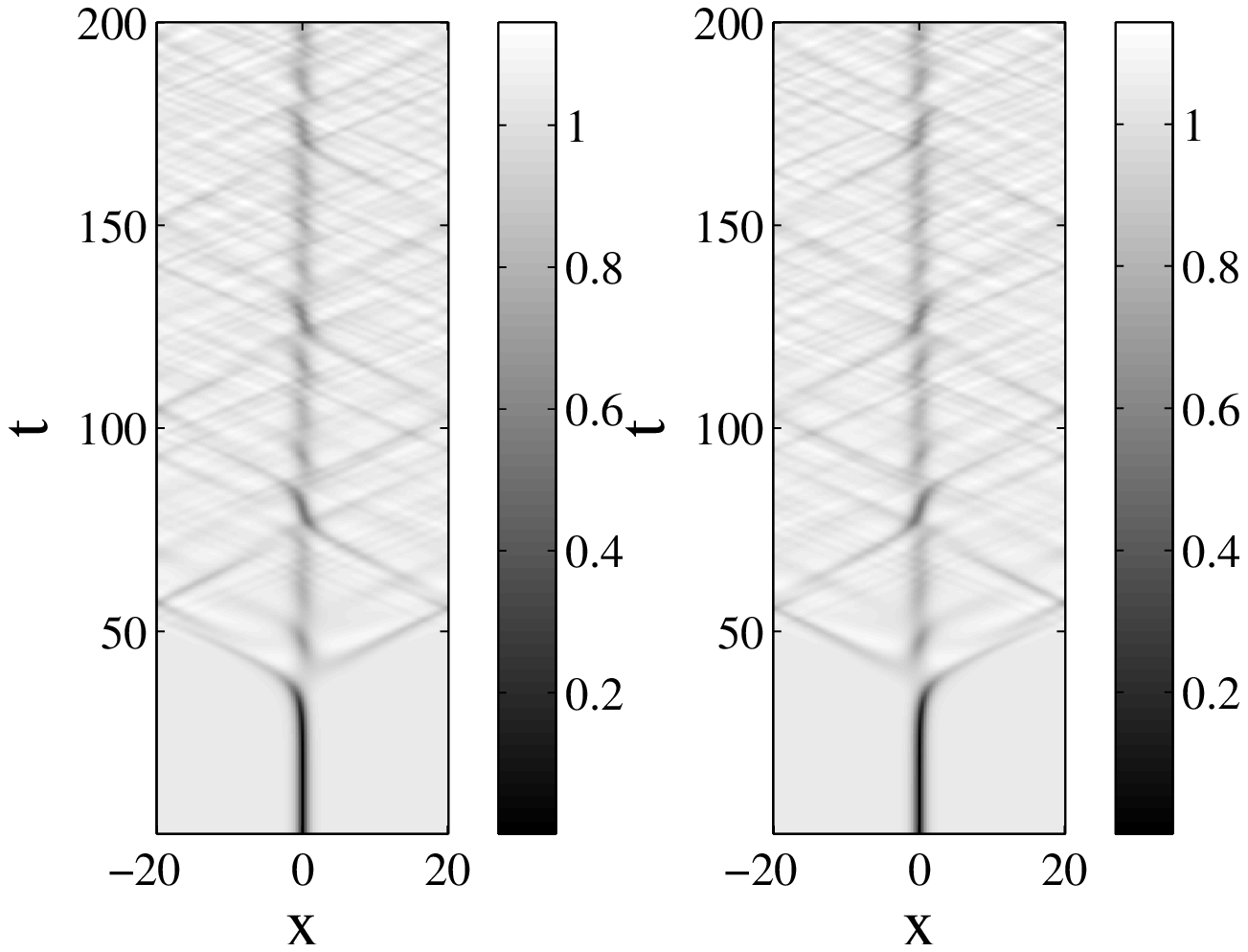}\label{evol}}
\caption{(a) Numerically obtained coupled dark solitons for $\omega=1$, $\mu=1$, and $k=0.1$. The black curves represent the real parts while the horizontal red lines are the imaginary parts of $\psi_j$, respectively. (b) The eigenvalues $\lambda$ of the solutions in (a) in the complex plane showing the instability of the solutions. (c) The critical eigenvalue of coupled dark solitons as a function of $k$. (d) An evolution of the unstable coupled dark solitons in (a). Shown is $|\psi_j|$. A grey-soliton-like structure in the centre at the end of the computation is an FA as the spatial profile of the phase difference between $\psi_1$ and $\psi_2$ form a $2\pi$-kink shape (not shown here).}
\end{center}
\end{figure*}

\begin{figure}[htbp!]
\begin{center}
\subfigure[]{\includegraphics[width=7cm]{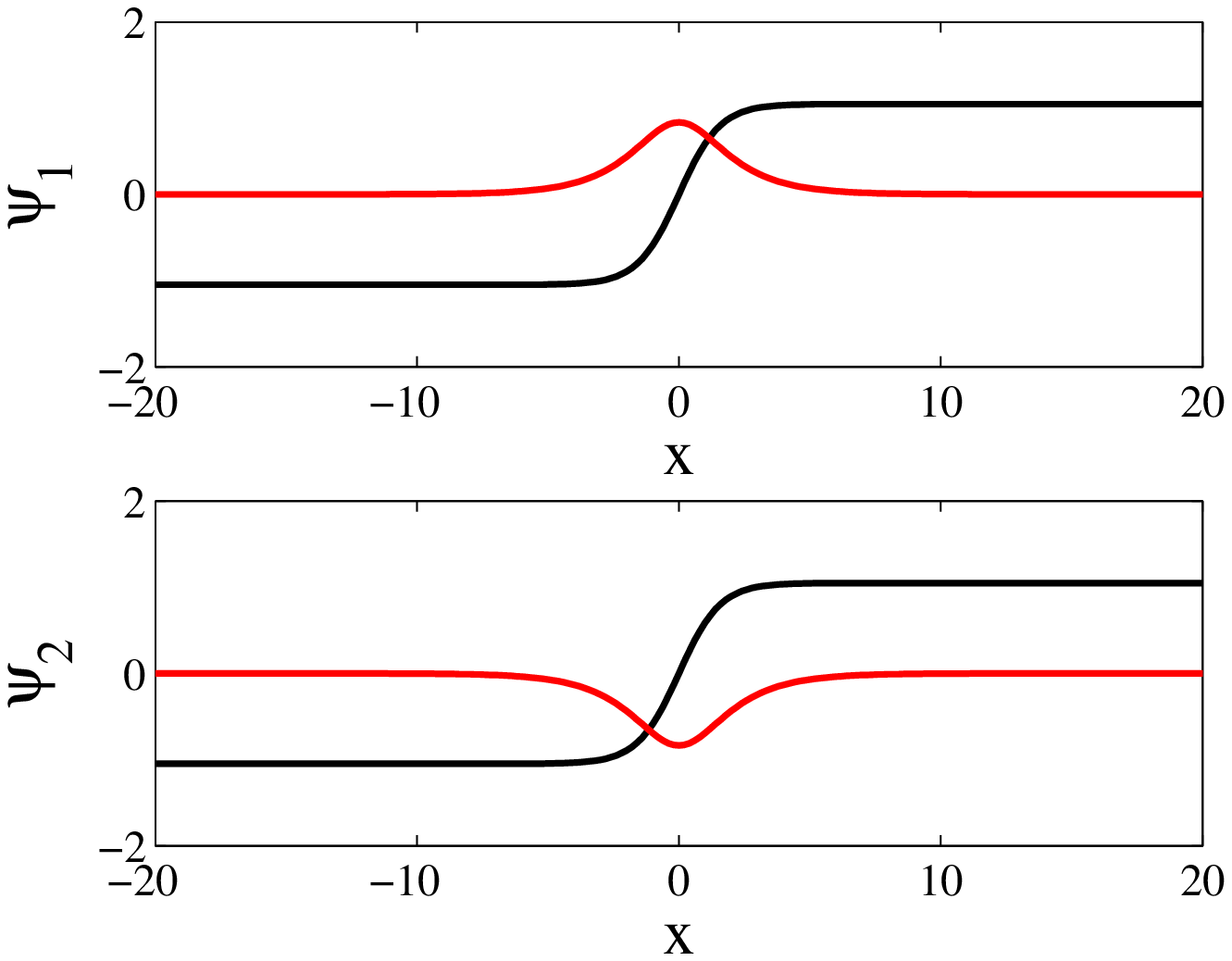}\label{Fig 5}}
\subfigure[]{\includegraphics[width=7cm]{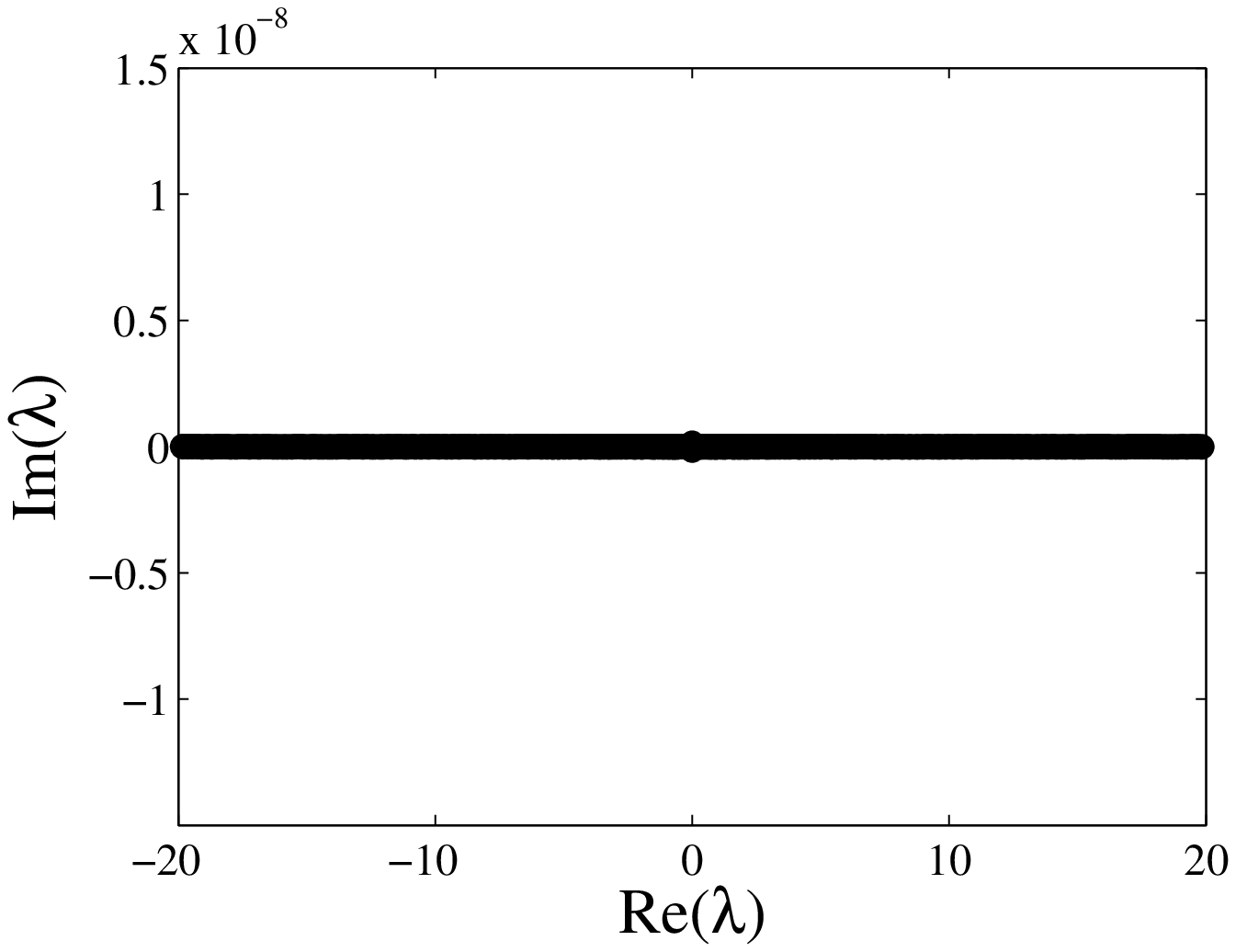}\label{Fig 3}}
\caption{(a) A numerically obtained FAs for $\omega=1$, $\mu=1$, and $k=0.1$. The black curves represent the real parts while the red curves are the imaginary parts of $\psi_1$ and $\psi_2$, respectively. (b) The eigenvalue distribution of the FAs in (a) in the complex plane showing the stability of the solution.}
\end{center}
\end{figure}

\subsection{The case $k\sim k_{ce}$ and $V\neq0$}

When there is a harmonic potential, the dynamics of the solutions will be determined perturbatively. We will particularly follow the method of \cite{kivs95,fran02,hong09}, which was developed for dark solitons, to our case.  A similar calculation will be performed to discuss the dynamics and stability of FAs in the limit of $k$ close to $k_{ce}$ as in that case the real part, i.e.\ the dark soliton component, is dominant and $C_1,\,C_2\approx0$.

Due to the presence of the magnetic trap, which is assumed to be slowly varying, i.e.\ $\Omega^2\ll1$, we take the ansatz
\begin{equation}
\Psi_j=\Psi_{TF}\Phi_j,
\label{TF}
\end{equation}
where $\Psi_{TF}$ is the Thomas-Fermi cloud approximately given by
\begin{equation}
\Psi_{TF}=\sqrt{\textrm{max}\{(1-V/(1+k)),0\}}.
\end{equation}
and
\begin{eqnarray}
\Phi_j&=&A\tanh\left(z\right)+i\left(C_j\sech\left(z\right)+\tilde v\right),\,j=1,2,
\label{psi0}
\end{eqnarray}
where $z=B\left(\tilde{x}-x_0\right)$. The parameters $A$, $B$, $C$, $x_0$ are in general functions of $\tilde{t}$. Here, $A^2+\tilde v^2=1$. When $C_j\equiv0$, one can recognise that the above function is the usual ansatz to describe a dark soliton travelling with the velocity $\tilde v$, where in that case the parameters $A,\,B,\,\tilde v$ and $x_0$ can be conveniently written as $A=B=\cos\varphi,$ $\tilde v=\sin\varphi$, and $x_0(\tilde t)=\int_0^{\tilde t}\sin\varphi(t')\,dt'$ \cite{fran02}.

Note the similarity of (\ref{TF}) with the ansatz $\Psi_j=\Psi_{TF}\psi_{ds}$ \cite{fran02,hong09} used to study the dynamics of dark solitons in a harmonic potential. Substituting the ansatz (\ref{TF}) into (\ref{nls3}) yields the equation for $\Phi_j$, i.e.\
\begin{eqnarray}
\displaystyle  i{\Phi_j}_{\tilde{t}}+\frac{1}{2}{\Phi_j}_{\tilde{x}\tilde{x}}-|\Phi_j|^2\Phi_j+\frac{\Phi_j+k\Phi_{3-j}}{1+k}\approx R(\Phi_j), \label{nls4}
\end{eqnarray}
where
\[R(\Phi_j)=\frac1{(1+k)}\left[\left(1-|\Phi_j|^2\right)V\Phi_j+\frac12V_{\tilde{x}}{\Phi_j}_{\tilde{x}}\right].\]
Here, for $R$ we only retain terms linear in $V$ and $V_{\tilde{x}}$, but neglecting terms linear in $V_{\tilde{x}\tilde{x}}$.

When there is no magnetic trap, (\ref{nls4}) with $V\equiv0$ can be derived from the Lagrangian
\begin{eqnarray}
\mathcal{L}&=&\int_{-\infty}^\infty\frac{1}2\sum_{j=1}^2i\left(\Phi_j^*{\Phi_j}_{\tilde{t}}-
\Phi_j{\Phi_j^*}_{\tilde{t}}\right)\left(1-\frac{1}{|\Phi_j|^2}\right)-|{\Phi_j}_{\tilde{x}}|^2-\left(|{\Phi_j}|^4-1\right)\nonumber\\
&&+2\frac{|{\Phi_j}|^2+k{\Phi_j}{\Phi_{3-j}^*}-(1+k)}{1+k}\,d\tilde{x} .
\label{lagv0}
\end{eqnarray}
It is not clear how to evaluate the first term of the integral (\ref{lagv0}) due to the present of nonzero $C_j$ in the denominator of $\left(1-{1}/{|\Phi_j|^2}\right)$. Because of that, in the following we assume $C_j$ to be small, i.e.\ we consider the case of $k\sim k_{ce}$, and take the series expansion with respect to $C_j$ about $C_j=0$ upto $\mathcal{O}(C_j^5)$. Performing the integration will yield the effective Lagrangian
\begin{eqnarray}
&&\mathcal{L}_{eff}={x_0}_{\tilde{t}}\biggr[-4A\tilde v+4\tan^{-1}({\frac{A}{\tilde v}})-\pi AC_3\biggr]-\frac{4}{3B}A^2(A^2+B^2)\nonumber\\
&&+\frac{\pi \tilde v}{B}(A^2C_3-C_5)-\frac{1}{3B}\biggr[(B^2-16A^2)C_4+2C_6\biggr]\nonumber \\
&&-\frac{2}{B(1+k)}\biggr[(3k+2)C_4-2kC_1C_2\biggr]+\mathcal{O}(C_j^5),
\label{leff}
\end{eqnarray}
where $C_{2+k}=(C_1)^k+(C_2)^k$, $k=1,\dots,4$. One can check that when the soliton is not moving, i.e.\ ${x_0}_{\tilde{t}}=\tilde v=0$ and $A=1$, the Euler-Lagrange equations for the remaining parameters $B,\,C_1,$ and $C_2$ 
can be solved analytically to yield
\begin{equation}
B^{(0)}=\frac{2\sqrt{k}}{\sqrt{1+k}},\,C_1^{(0)}=-C_2^{(0)}=\pm\frac{\sqrt{1-3k}}{\sqrt{1+k}},
\label{bc}
\end{equation}
which is nothing else, but the solution (\ref{jv}) or $B^{(0)}=1,\,C_j^{(0)}=0$, which is the dark soliton (\ref{ds}). It is important to note here that despite the series expansion with respect to $C_j$ that we took prior to integrating (\ref{lagv0}), our result (\ref{bc}) corresponds to an exact solution for any value of $0\leq k\leq k_{ce}$. It is not yet clear to us why this is the case.

To determine the influence of the magnetic trap, i.e.\ $V\neq0$, to the FAs, we will follow the idea of \cite{kivs95} and treat the right hand side of (\ref{nls4}) as perturbations. Due to the presence of the perturbations the Euler-Lagrange equation for the variable $\alpha_j$, with $\alpha_j=\varphi,\,B,$ and $C_j$, becomes \cite{kivs95}
\[
\frac{\partial \mathcal{L}_{eff}}{\partial{\alpha_j}}-\partial_{\tilde{t}}\frac{\partial\mathcal{L}_{eff}}{\partial{{\alpha_j}_{\tilde{t}}}}=2\,\textrm{Re}\left(\int_{-\infty}^\infty \sum_{j=1}^2R(\Phi_j)^*\frac{\partial\Phi_j}{\partial{\alpha_j}}\,d\tilde{x}\right),
\]
Following \cite{fran02,fran10}, we need two active variables only. Because of that, we will assume that adiabatically $B$ and $C_j$ are independent of time and are given by (\ref{bc}). Hence, we obtain the following set of equations	
\begin{eqnarray}
\frac{d A}{d\tilde{t}}&=& \frac{\Omega^2 x_0 \tilde{v}}{12\sqrt{k}(k+1)^\frac{5}{2}} \biggr[A^2(k+1)+11k-1\biggr]\label{lageq4},\\
\frac{d x_0}{d\tilde{t}}&=& \frac{\tilde{v}}{576A k^\frac{3}{2}(k+1)^\frac{5}{2}} \biggr[192k^4(A^2+14)+576k^3(A^2+8) \nonumber \\
&+&576k^2(A^2+2)+\Omega^2 k^2 \biggr\{(48{x_0}^2+\pi^2)(A^2+3) \nonumber \\
&+&12(A^2+9)\biggr\}+2k\Omega^2\biggr\{24{x_0}^2(A^2-1)+12(A^2+4) \nonumber \\
&+&\pi^2(A^2+1)\biggr\}+\Omega^2(A^2-1)(\pi^2+12) +192k(A^2-4)\biggr].\label{lageq5}
\end{eqnarray}
From the two equations above, we obtain (see, e.g., \cite{hong09})
\begin{eqnarray}
\frac{d^2x_0}{d{t}^2}=\frac{(1-5k)\Omega^2}{1+k}x_0+\mathcal{O}(\Omega^4),
\label{x0}
\end{eqnarray}
where we have assumed that $A\approx1$ and $x_0\approx0$. Note that the derivative in (\ref{x0}) is with respect to the original time variable. This shows that the solution is stable for values of $k>\frac{1}{5}$. When $k=1/3$, we recover the oscillation frequency of dark solitons in a harmonic trap \cite{busc00} (see also \cite{fran02,fran10,theo10}). We will check the validity of our approximation by comparing it with numerical results.

\begin{figure}[htbp!]
\begin{center}
\subfigure[]{\includegraphics[width=7cm]{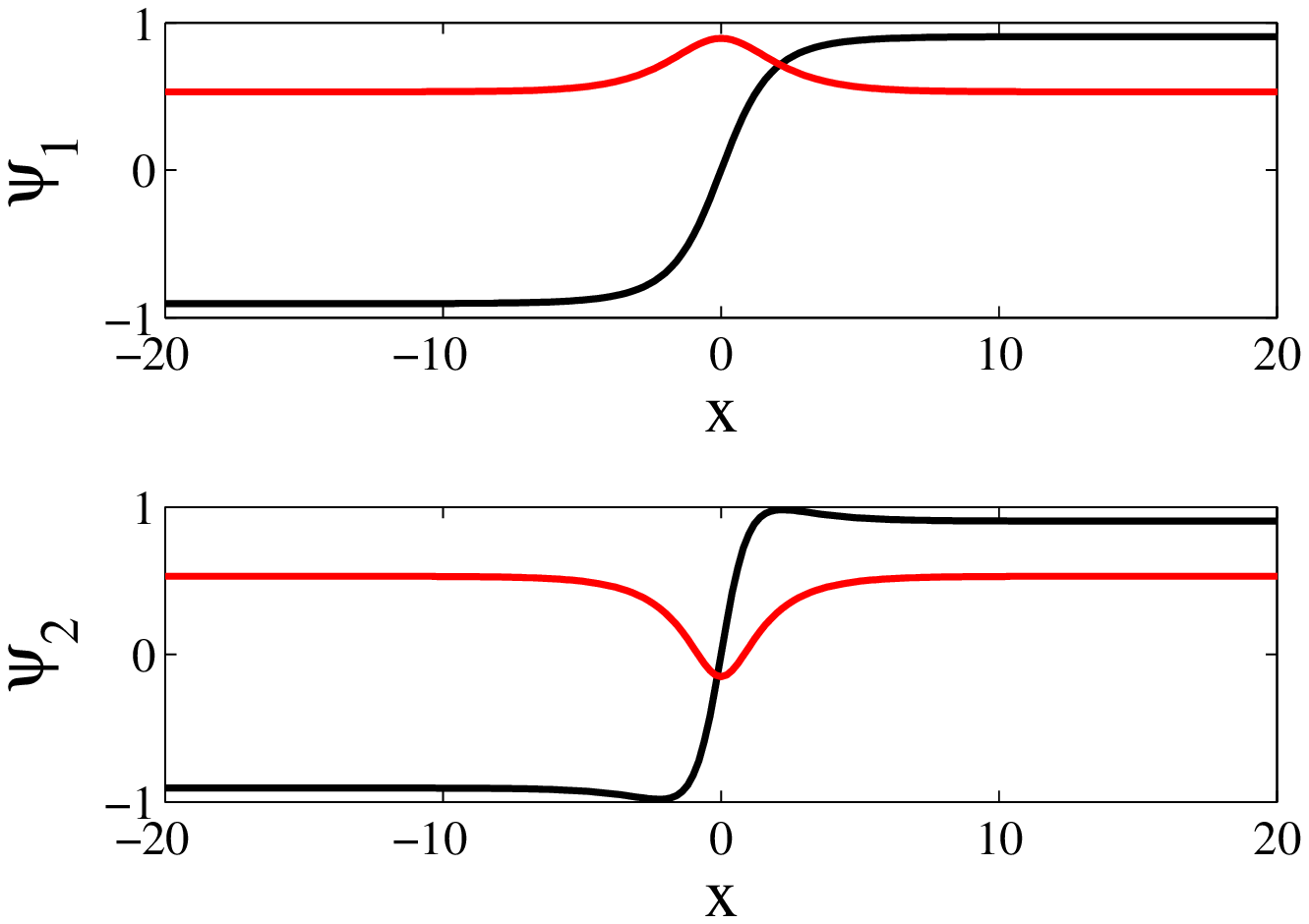}\label{Fig.7a}}
\subfigure[]{\includegraphics[width=7cm]{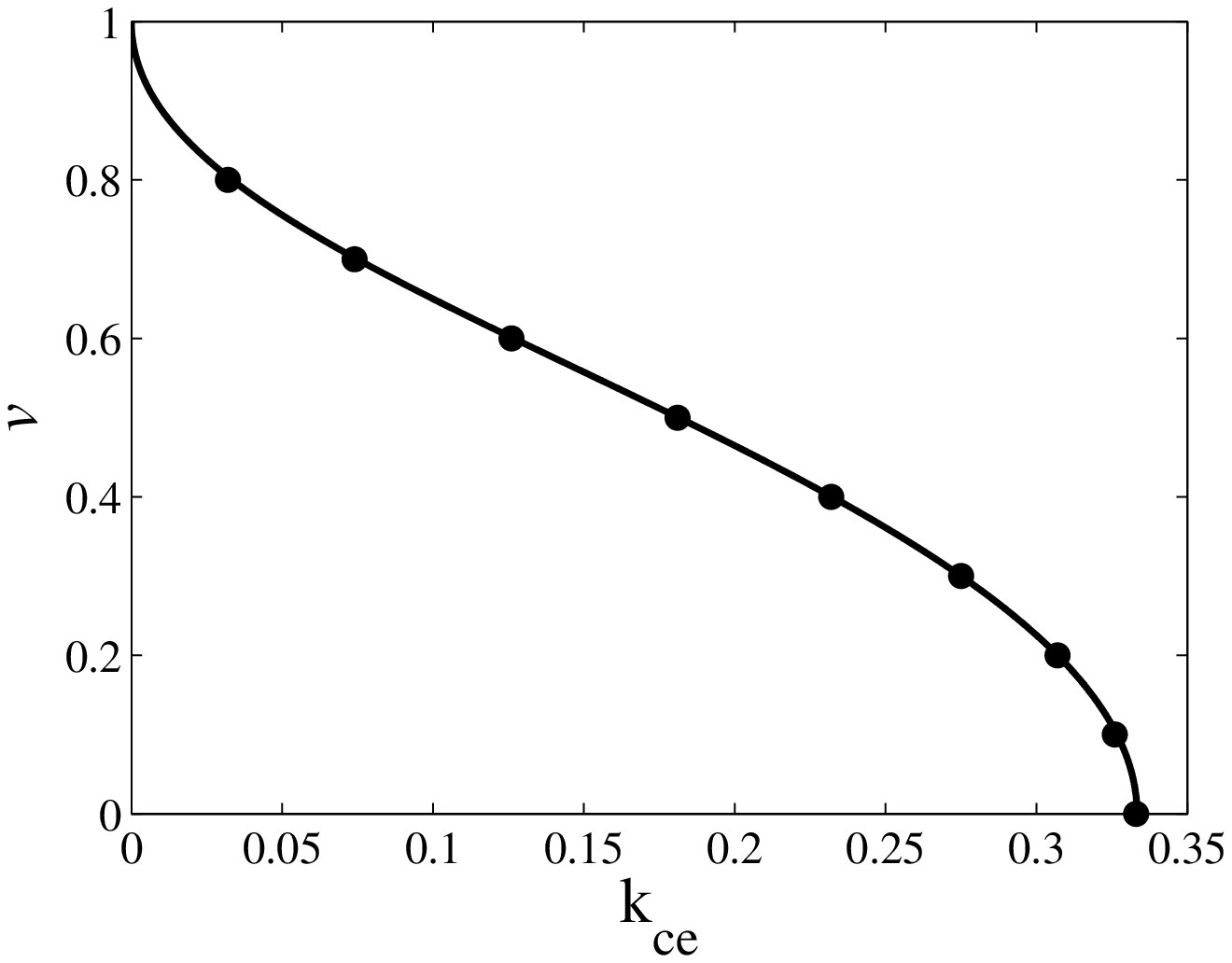}\label{Fig.6}}
\end{center}
\caption{(a) A numerically obtained FAs for 
$v=0.2$ and $k=0.1$. 
(b) The critical coupling constant $k_{ce}$ as a function of the velocity $v$ for the existence of travelling FAs. Filled circles are numerical data and the solid line is the function in (\ref{k_v}). FAs only exist on the left of the solid curve.}
\end{figure}

\subsection{The case $k\sim0$ and $V\neq0$}

When $k$ is close to zero, the imaginary part of the FAs is dominant over the real part. Due to the presence of a magnetic trap, the imaginary part will become the Thomas-Fermi ground state in the limit $k\to0$. Motivated by, e.g., \cite{kivs01} where a Gaussian ansatz, which is the ground state solution of the linear Schr\"odinger equation (\ref{nls2}) and (\ref{ep}) with $\mu=k=0$, was shown to be able to approximate the ground state of the nonlinear equation, in the following we will take a similar ansatz for our problem. For the present case, we will consider the governing equation (\ref{nls2}) to be derived from the Lagrangian
\begin{eqnarray}
\mathcal{L}&=&\int_{-\infty}^\infty\frac{1}2\sum_{j=1}^2i\left(\psi_j^*{\psi_j}_{{t}}-\psi_j{\psi_j^*}_{{t}}\right)-|{\psi_j}_{{x}}|^2-\mu|{\psi_j}|^4\nonumber\\&&+2\left(\omega-V\right)|{\psi_j}|^2+2k{\psi_j}{\psi_{3-j}^*}\,\,d{x}.
\label{lag1}
\end{eqnarray}
Note that the main difference between (\ref{lag1}) and (\ref{lagv0}) is the factor $\left(1-\frac{1}{|\Psi_j|^2}\right)$ in the first term of the integrand. As for the ansatz, we take
\begin{equation}
\psi_j=\left(A_jx+iC_j\right)e^{-Bx^2}.
\label{lagk0}
\end{equation}
One advantage of such a Gaussian ansatz is that the Lagrangian can be evaluated rather straightforwardly. This advantage is expected to be particularly useful in studying bound states of FAs.

Substituting the ansatz into the Lagrangian (\ref{lag1}) yields the effective Lagrangian
\begin{eqnarray}
\mathcal{L}_{eff}&=&-\frac{\sqrt{\pi}}{256B^{\frac{5}{2}}}\biggr[4\sqrt{2} A_3(12B^2+3\Omega^2-8\omega B) \nonumber \\
&+&16\sqrt{2}BC_4(4B^2+\Omega^2-8\omega B)+64B^2C_6 \nonumber \\
&+&16B(A_6-4\sqrt{2}kA_7)+3A_4\biggr],\label{leff2}
\end{eqnarray}
from which we obtain the Euler-Lagrange equations
\begin{eqnarray}
&&-16\sqrt{2}kA_{3-j}B+A_j\biggr[3A_j^2+6\sqrt{2}\Omega^2+8\sqrt{2}B(3B-2\omega+\frac{{C_j}^2}{\sqrt{2}})\biggr]=0,\label{lageq7} \\
&&8\sqrt{2}BkC_{3-j}+C_j\biggr[{A_j}^2+\sqrt{2}\Omega^2 +4\sqrt{2}B(B-2\omega+\sqrt{2}{C_j}^2)\biggr]=0,\label{lageq9}\\
&&15A_4+16B(3A_6+4BC_6)+16\sqrt{2}BC_4\left(3\Omega^2-4B^2-8B\omega\right) \nonumber \\
&&+12\sqrt{2}A_3\left(5\Omega^2+4B^2-8B\omega\right) -64\sqrt{2}kB\left(4BC_1C_2+3A_1A_2\right)=0,\label{lageq11}
\end{eqnarray}
where $A_3=A_1^2+A_2^2$, $A_4=A_1^4+A_2^4$, $A_5=A_1C_1+A_2C_2$, $A_6=A_1^2C_1^2+A_2^2C_2^2$, $A_7=A_1A_2+4BC_1C_2$.
Solving the system of algebraic equations above yields an approximation to the FAs. The validity of the ansatz will be checked by comparing the results presented here with numerical results. The comparison is presented in section \ref{IIIB}. As for the stability, using a Gaussian ansatz, one will need to include, e.g., chirp variables, which we leave for future investigation.




\section{Numerical computations}

To study the existence of static FAs in the time-independent framework of (\ref{nls2}), we use a Newton-Raphson continuation method with an initial ansatz (\ref{jv}) or (\ref{ds}). The spatial first and second order derivatives (the former being in the governing equation in a moving coordinate frame) are approximated using central finite difference with three-point or five-point stencils. At the computational boundaries, we use the Neumann boundary conditions. Numerical linear stability analysis of a solution $\psi^{(0)}_j(x)$ is then performed by looking for perturbed solutions of the form
\[
\psi_j=\psi^{(0)}_j(x)+\epsilon[a_j(x)\,e^{i\lambda t}+b_j^*(x)\,e^{-i\lambda^* t}],\,j=1,2.
\]
Substituting the ansatz into the governing equation (\ref{nls2}) and keeping the linear terms in $\epsilon$, one will obtain a linear eigenvalue problem for the stability of $\psi^{(0)}_j$. The ensuing eigenvalue problem is then discretized using a similar finite difference scheme as above and solved numerically for the eigenfrequency $\lambda$ and corresponding eigenfunctions $a_j$ and $b_j$. It is then clear that $\psi^{(0)}_j(x)$ is a stable solution if the imaginary parts of all the eigenvalues vanish, i.e.\ $\textrm{Im}(\lambda)=0$.

The evolutions in time of the solitons when they are unstable are investigated through integrating the time-dependent governing equations using a Runge-Kutta method of order four.

\subsection{Coupled BECs without a trap: $\Omega=0$}

As pointed out in \cite{kaur05,kaur06} for $k$ small enough, coupled dark solitons (\ref{ds}) are unstable. Shown in Figs.\ \ref{Fig 1} and \ref{Fig 4} are respectively numerically obtained coupled solitons and their eigenvalue structure in the complex plane for $k=0.1$. Because there is at least one pair of eigenvalues with nonzero imaginary parts, we conclude that the solitons are unstable. We have calculated the stability of dark solitons for different values of coupling constant $k$, where we obtained that dark soliton solutions are stable for $k\geq{1}/{3}$ as shown in Fig.\ \ref{Fig 2}. In the instability region of the coupled dark solitons, one can obtain stable FAs \cite{kaur05,kaur06}. Shown in Fig.\ \ref{evol} is the typical time-dynamics of unstable coupled dark solitons, where an FA is obtained spontaneously. Nonetheless, it is important to note that it is not always the case. When the coupling constant is too small, instead of yielding FAs, the dark solitons will repel and move away from each other, which is not shown here. Hence, coupled dark solitons will transform into FAs if the coupling is in an intermediate region.

We depict in Figs.\ \ref{Fig 5} and \ref{Fig 3} a numerically obtained FAs and its eigenvalues, respectively. We observed that the FAs solutions indeed exist stably only for $k<1/3$. As $k$ approaches $1/3$, the amplitude of the humps in the imaginary parts tends to zero and the FAs changes into coupled dark solitons.

Because solitons of (\ref{nls2}) without a trap are translationally invariant, they can move freely in space. This motivates us to study the existence and stability of solitons in a moving coordinate frame $\xi=x-vt$ (cf.\ (\ref{nls-3})).

We present in Fig.\ \ref{Fig.7a} an FA travelling with $v=0.2$ for $k=0.1$. One can see the deformation in the shape of the soliton due to the nonzero value of $v$. The shape of the deformation suggests the correction ansatz (\ref{corr}). For a fixed $v$, if the coupling constant is increased further, the FAs changes into coupled dark solitons at a critical value $k_{ce}<1/3$. The plot of $k_{ce}$ as a function of $v$ is shown in Fig.\ \ref{Fig.6} as filled circles. The solid curve in the same figure is the graph of (\ref{k_v}), where we obtain perfect agreement. We do not show the profile of dark solitons for nonzero $v$ as exact solutions are available.


After examining the existence of the solitons, next we study their stability. We have calculated the stability of FAs for several values of $k$ and find that the FAs are always stable in their existence domain. Extending the calculation to coupled dark solitons, we found that they are unstable for the values of the coupling constant $k< k_{ce}$ corresponding to each value of the velocity $v$. For the values of $k\geq k_{ce}$, the dark soliton becomes stable. The Runge-Kutta method of order four has been used to verify the time dynamics of the FAs as well as dark soliton.

\begin{figure*}[htbp!]
\begin{center}
\subfigure[]{\includegraphics[width=7cm]{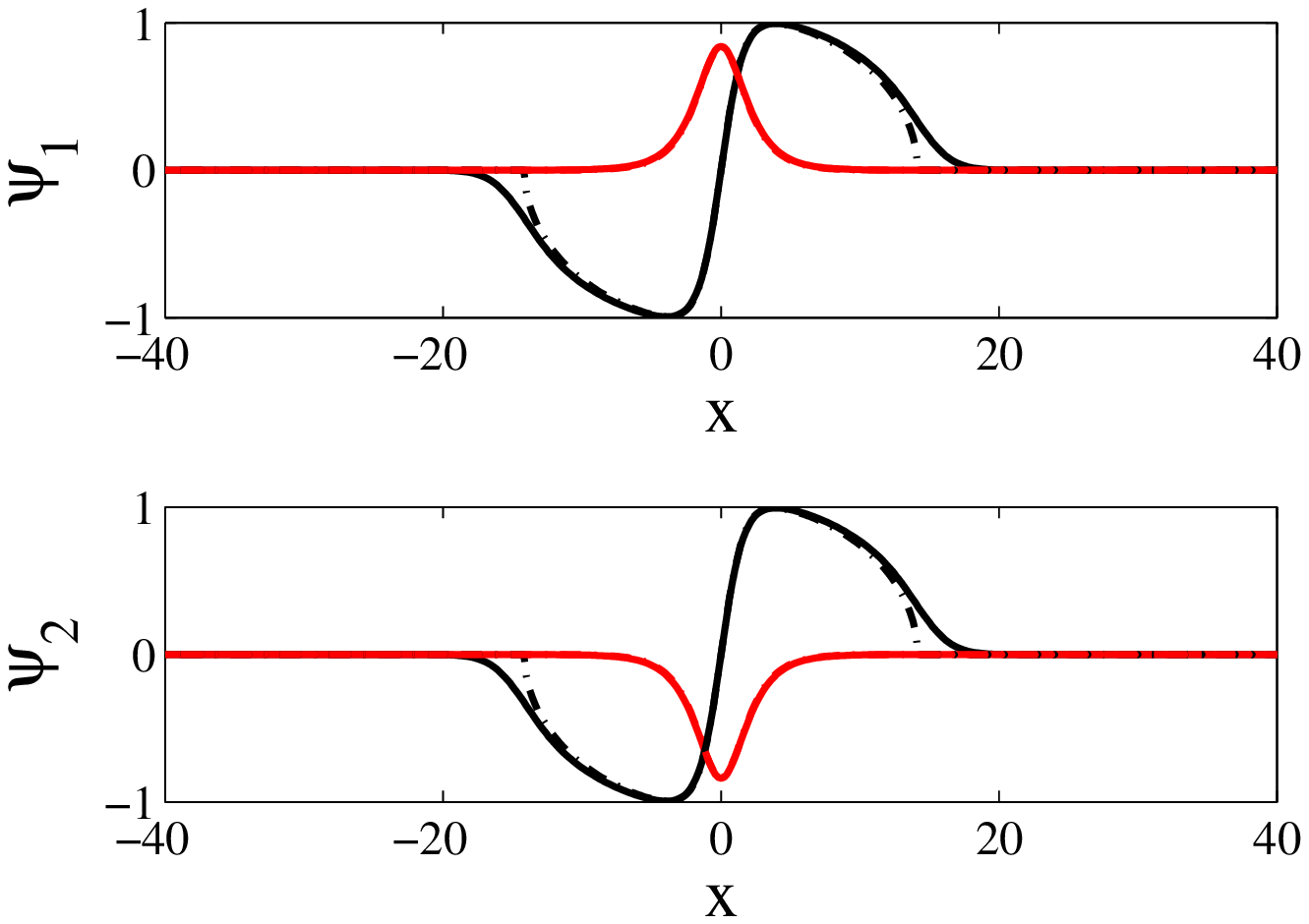}\label{Fig.12}}
\subfigure[]{\includegraphics[width=7cm]{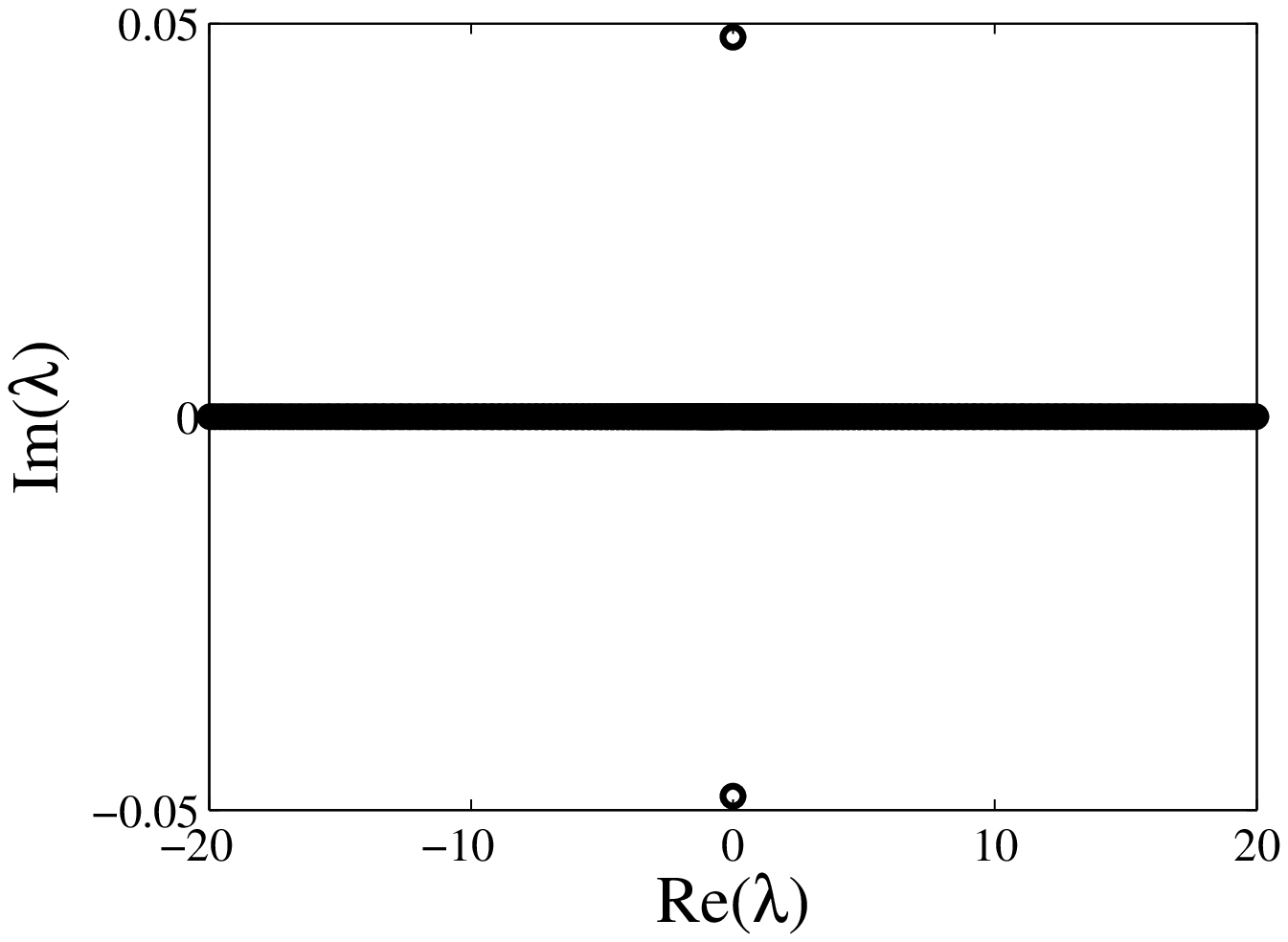}\label{Fig.15}}
\subfigure[]{\includegraphics[width=7cm]{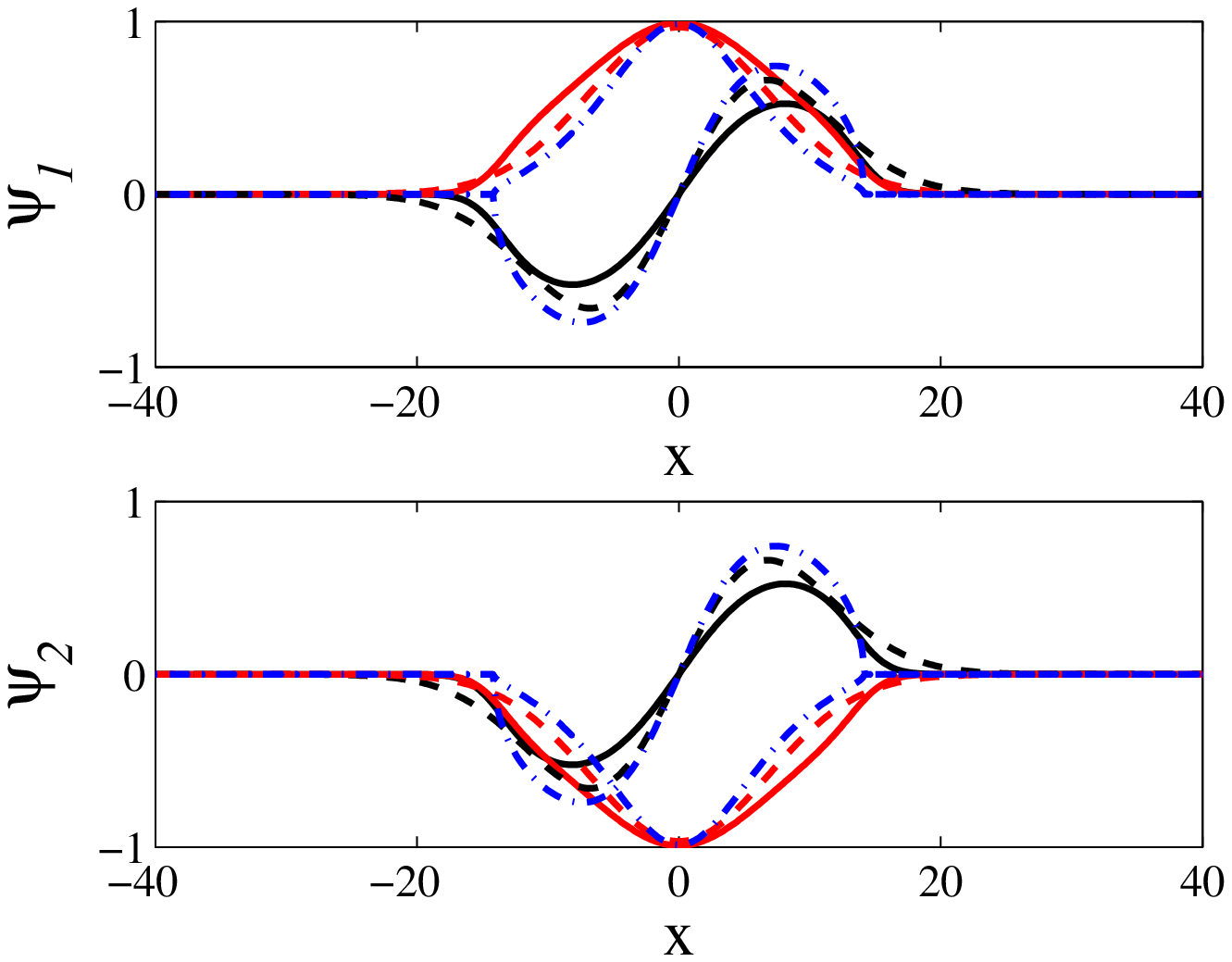}\label{Fig.va212}}
\subfigure[]{\includegraphics[width=7cm]{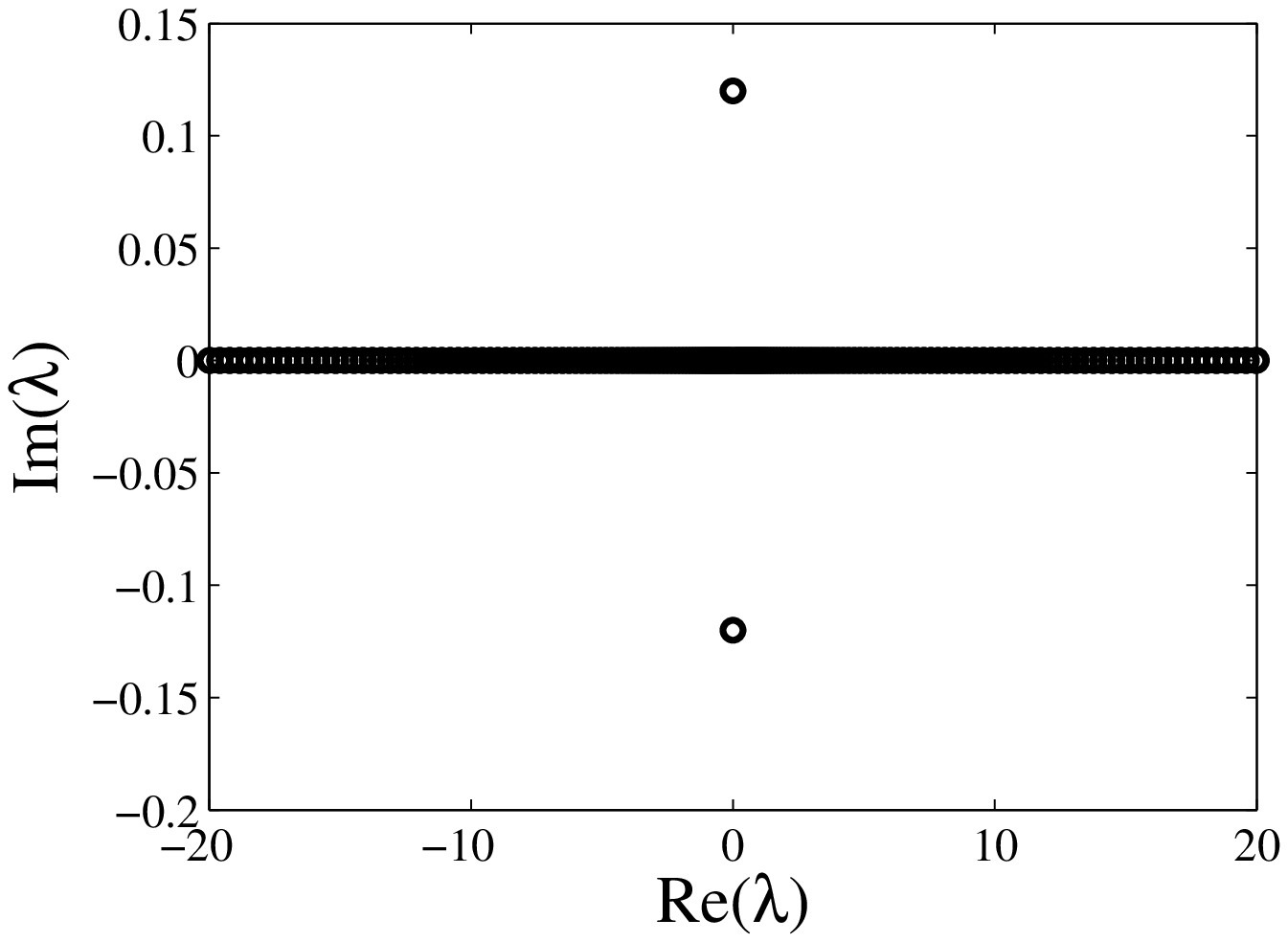}\label{Fig.14}}
\end{center}
\caption{Numerically obtained FAs solutions of (\ref{nls2}) and their eigenvalues for $\Omega=0.1$, $\omega=1$, $\mu=1$, (a-b) $k=0.1$, (c-d) $k=0.008$. The dash-dotted curves in (a)-(c) are the Thomas-Fermi ansatz (\ref{TF}) and the dashed curves in (c) are approximations obtained through the variational approach using the Gaussian ansatz (\ref{lagk0}).}
\end{figure*}

\begin{figure}[htbp!]
\begin{center}
\includegraphics[width=10cm]{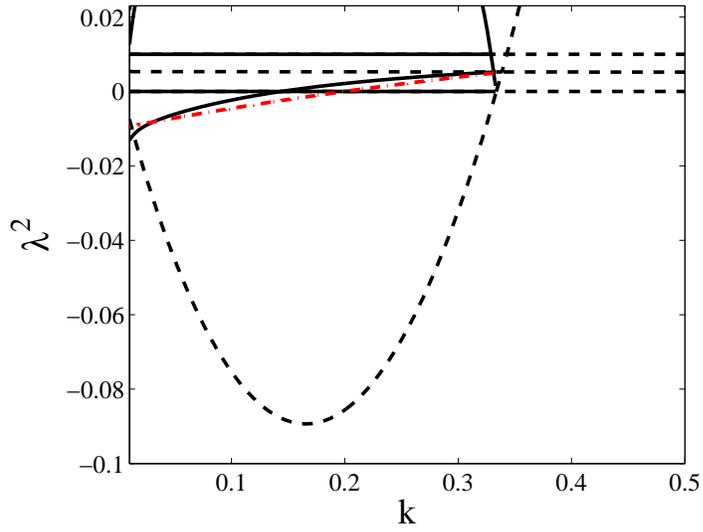}
\end{center}
\caption{The first few lowest squared eigenvalues of FAs (solid) and dark solitons (dashed) as a function of $k$ for $\Omega=0.1$. The dash-dotted curve represents the approximation (\ref{x0}). Shown is $|\psi_j|$. }
\label{Fig16jvlowev}
\end{figure}

\subsection{Coupled BECs with a trap: $\Omega\neq0$}\label{IIIB}

\begin{figure}[htbp!]
\begin{center}
\includegraphics[width=7cm]{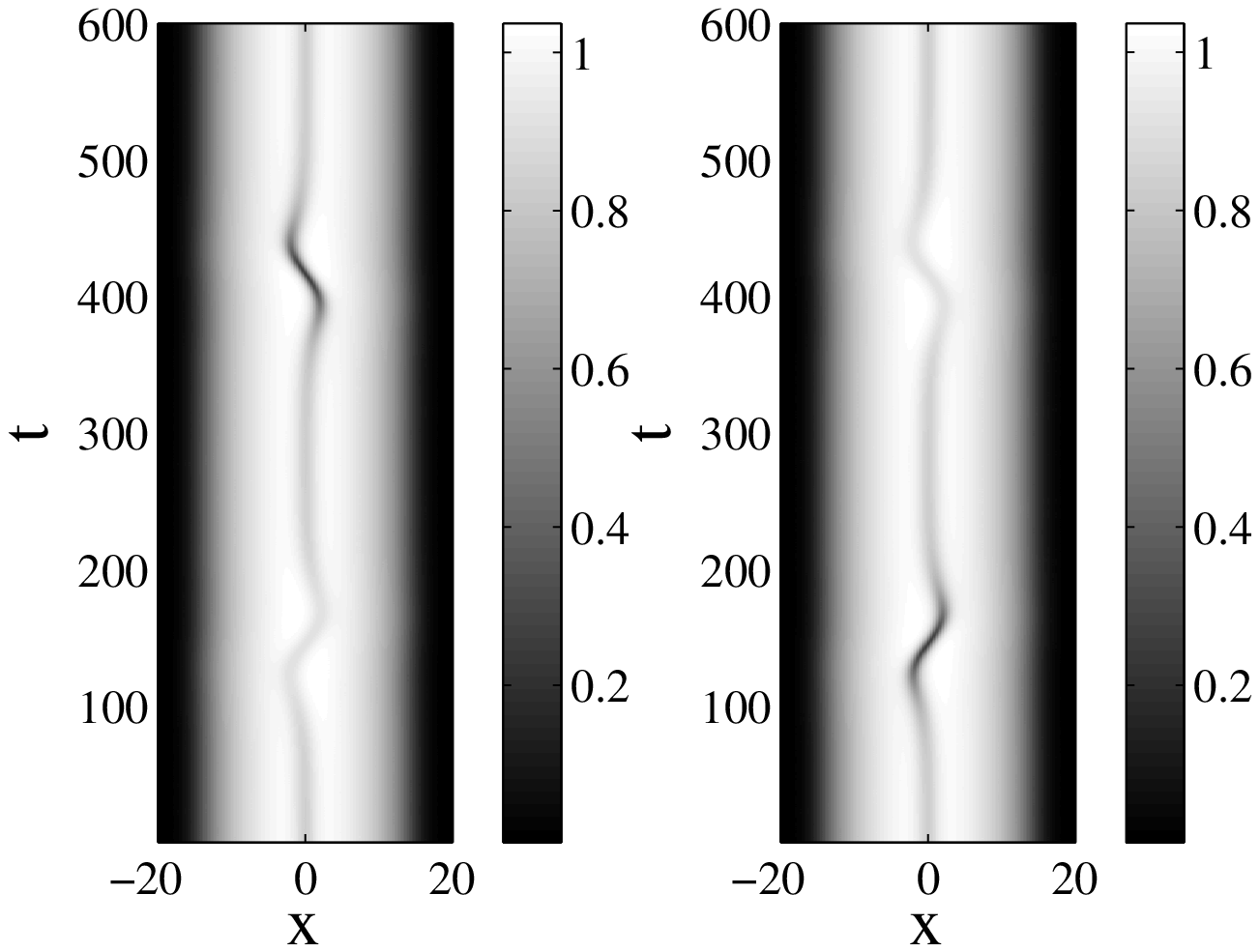}
\caption{A numerical evolution of the FAs in Fig.\ \ref{Fig.12} for $\Omega=0.1$ and $k=0.1$. Shown is $|\psi_j|$.}
\label{Fig.17}
\end{center}
\end{figure}

Next, we consider the existence and stability of FAs and coupled dark solitons in the presence of a harmonic trap. Two FAs for two different values of coupling constant $k$ are shown in Figs.\ \ref{Fig.12} and \ref{Fig.va212} with $\Omega=0.1$. In both panels, we present in dash-dotted curves our approximation (\ref{TF}). In panel (c), we also depict our Gaussian approximation (\ref{lagk0}), where one can note that the ansatz is slightly better than the Thomas-Fermi ansatz (\ref{TF}). The relatively good agreement deviates rapidly as $k$ increases towards the critical coupling. In Figs.\ \ref{Fig.15} and \ref{Fig.14}, we present the eigenvalues of the FAs, where the solitons are found to be unstable and this agrees with the result obtained in (\ref{x0}).

The finding that the FAs above are unstable suggests us to investigate whether FAs are always unstable for nonzero $\Omega$. The result is summarized in Fig.\ \ref{Fig16jvlowev}. Shown in solid curves are the first few lowest squared eigenvalues as a function of $k$ for $\Omega=0.1$. As stability corresponds to $\lambda^2>0$, the figure shows that there is a critical value $k_{cs}$ above which FAs are stable, which for the parameter value above is approximately 0.145. We also present the analytical result (\ref{x0}) in dashed-dotted line, where qualitatively good agreement is obtained. 

\begin{figure}[htbp!]
\begin{center}
\includegraphics[width=7cm]{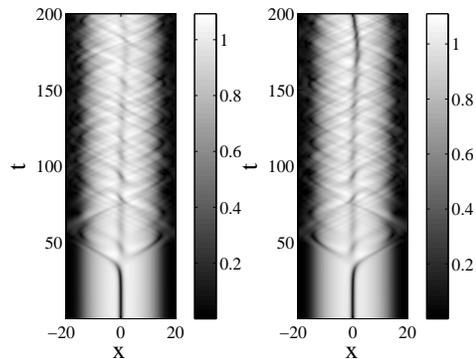}
\caption{Numerical evolution of coupled dark solitons for $\Omega=0.1$ and $k=0.1$.}
\label{Fig.25}
\end{center}
\end{figure}

We also perform the time dynamics of the unstable FAs in Fig.\ \ref{Fig.12}. Shown in Fig.\ \ref{Fig.17} is the typical evolution of unstable FAs, where one can see that the instability causes the solitons to oscillate about the minimum of the trap non-sinusoidally. Hence, despite the instability, the FAs still persists. This result rather applies generally to other values of coupling constants where FAs are unstable.

As for the existence of FAs in the presence of a magnetic trap, we observe that the critical coupling $k_{ce}$ above which FAs do not exist is almost independent of the trapping parameter $\Omega$, which is reasonable as the width of the non-zero imaginary part of the FAs solution is small compared to the width of the trapping parameter $\Omega$ when $k\to k_{ce}$.

We have also studied the existence and the stability of dark solitons. Shown as dashed lines in Fig.\ \ref{Fig16jvlowev} are the first few lowest squared eigenvalues of coupled dark solitons, where similarly as before for $k$ less than the critical coupling $k_{ce}$ coupled dark solitons may transform into an FA as shown in Fig.\ \ref{Fig.25}.

\section{Conclusion}

We have studied the existence and the stability of FAs and coupled dark solitons in linearly coupled Bose-Einstein condensates, both in the absence or presence of a harmonic potential. Without a trap, we have obtained travelling FAs and determined the range of values of the coupling constant $k$ where they exist. In the presence of a magnetic trap, it has been demonstrated that one can obtain a stable FAs by controlling $k$. In the region of $k$ where an FA is unstable, it has been shown that the FAs still exists, but oscillates about the minimum of the trap non-sinusoidally. 
Theoretical approximations based on variational formulations have been derived and used to explain the numerical results.

In the future, 
interaction of multiple FAs and dark solitons in the same setup as well as the existence and the stability of higher dimensional excitations, such as vortices, will also be investigated and reported elsewhere.

\ack

We thank the anonymous referees for their comments and suggestions that help improve the manuscript. M.I. Qadir acknowledges the University of Engineering and Technology, Lahore, Pakistan for financial support.


\section*{References}
\begin{thebibliography}{99}

\bibitem{jose62} B.D. Josephson, Phys.Lett. 1, 251 (1962).

\bibitem{smer97} A. Smerzi, S. Fantoni, S. Giovanazzi, and S. R. Shenoy,
Phys. Rev. Lett. 79, 4950 (1997).
\bibitem{ragh99} S. Raghavan, A. Smerzi, S. Fantoni, and S. R. Shenoy, Phys. Rev. A 59, 620-633 (1999).
\bibitem{giov00} S. Giovanazzi, A. Smerzi, and S. Fantoni,
Phys. Rev. Lett. 84, 4521 (2000).

\bibitem {albi05} M. Albiez, R. Gati, J. Folling, S. Hunsmann, M. Cristiani, and M.K. Oberthaler,
Phys. Rev. Lett. 95, 010402 (2005).

\bibitem{levy07} S. Levy, E. Lahoud, I. Shomroni, and J. Steinhauer,
Nature 449 579 (2007).

\bibitem {cata01} F.S. Cataliotti, S. Burger, C. Fort, P. Maddaloni, F. Minardi, A, Trombettoni, A. Smerzi, and M. Inguscio,
Science 293, 843 (2001).

\bibitem{ostr00} E. A. Ostrovskaya, Yu. S. Kivshar, M. Lisak, B. Hall, F. Cattani, and D. Anderson,
Phys.\ Rev.\ A 61, 031601-4 (2000).


\bibitem{anan06} D. Ananikian and T. Bergeman, Phys. Rev. A 73, 013604 (2006); Phys. Rev. A 74, 039905(E) (2006).

\bibitem{jia08} X.Y. Jia, WD. Li, and J.Q. Liang, Phys. Rev. A 78 023613 (2008).



\bibitem {baro82} A. Barone and G. Paterno, \emph{Physics and Applications of the Josephson Effect} (John Wiley \& Sons, New York, Singapore, 1982).

\bibitem {kaur05} V. M. Kaurov and A. B. Kuklov, 
Phys. Rev. A 71, 011601 (2005).

\bibitem {kaur06} V. M. Kaurov and A. B. Kuklov, 
Phys. Rev. A 73, 013627 (2006).

\bibitem {usti98} A. V. Ustinov, 
Physica D 123, 315 (1998).


\bibitem{bran} J. Brand, T.J. Haigh, and U. Z\"ulicke, Phys. Rev. A 80, 011602(R) (2009)




\bibitem{tril88} S. Trillo, S. Wabnitz, EM Wright, and GI Stegeman, 
Opt. Lett. 13, 672–674 (1988).

\bibitem{kivs89} Yu.S.\ Kivshar and B.A.\ Malomed, 
Opt. Lett. 14, 1365-1367 (1989).

\bibitem{akhm93} N.\ Akhmediev and A.\ Ankiewicz, 
Phys. Rev. Lett. 70, 2395–2398 (1993).

\bibitem{dror09} N.\ Dror and B.A.\ Malomed, Phys. Rev. E 79, 016605 (2009).

\bibitem{kono08} V.V.\ Konotop, 
in \emph{Emergent Nonlinear Phenomena in Bose-Einstein Condensates Theory and Experiment}, Eds.\ P.G.\ Kevrekidis, D.J.\ Frantzeskakis, R.\ Carretero-Gonz\'alez (Springer Berlin Heidelberg, 2008) pp.\ 65-97.

\bibitem{busc00}
Th. Busch and J. R. Anglin , 
Phys. Rev. Lett. 84, 2298–2301 (2000).

\bibitem{fran10} D. J. Frantzeskakis, 
J. Phys. A: Math. Theor. 43, 213001 (2010).

\bibitem{peli05} D.E. Pelinovsky, D. J. Frantzeskakis, and P. G. Kevrekidis, Phys. Rev. E 72, 016615 (2005). 




\bibitem{theo10} G. Theocharis, A. Weller, J. P. Ronzheimer, C. Gross, M. K. Oberthaler, P. G. Kevrekidis, and D. J. Frantzeskakis, Phys. Rev. A {\bf81}, 063604 (2010). 

\bibitem{well08} A. Weller, J. P. Ronzheimer, C. Gross, J. Esteve, M. K. Oberthaler, D. J. Frantzeskakis, G. Theocharis, and P. G. Kevrekidis, Phys. Rev. Lett. {\bf101}, 130401 (2008). 

\bibitem{burg99} 
S. Burger, K. Bongs, S. Dettmer, W. Ertmer, K. Sengstock, A. Sanpera, G. V. Shlyapnikov, and M. Lewenstein
Phys. Rev. Lett. 83, 5198 (1999).

\bibitem{beck08} C. Becker, S. Stellmer, P. Soltan-Panahi, S. D\"orscher, M.\ Baumert, E-M. Richter, J. Kronj\"ager, K. Bongs, and K. Sengstock,
    Nature Phys. 4, 496 (2008).

\bibitem{stel08} S. Stellmer, C. Becker, P. Soltan-Panahi, E.-M. Richter, S. D\"{o}rscher, M. Baumert, J. Kronj\"{a}ger, K. Bongs, and K. Sengstock
Phys.\ Rev.\ Lett.\ 101, 120406 (2008).


\bibitem{kivs95} Y.S.\ Kivshar and W.\ Krolikowski, Opt.\ Comm.\ 114, 353--362 (1995).

\bibitem{kivs98} Yu.S.\ Kivshar and B.\ Luther-Davies, Phys. Reports 298, 81-197 (1998). 



\bibitem{fran02} D.J. Frantzeskakis, G. Theocharis, F. K. Diakonos, P.\ Schmelcher, and Yu.S. Kivshar, Phys.\ Rev.\ A {\bf66}, 053608 (2002). 

\bibitem{hong09} L.\ Hong and W.\ Dong-Ning, 
    Chinese Phys. B 18, 2659 (2009).


\bibitem{kivs01} Kivshar Yu S, Alexander T J and Turitsyn S K 2001 Phys.\ Lett.\ A 278 225



\end{thebibliography}
\end{document}